\documentclass[twocolumn,aps,prb,superscriptaddress,showpacs]{revtex4-1} 
\usepackage{graphicx} 
\usepackage[version=3]{mhchem} 
\usepackage[colorlinks=true, citecolor=blue,linkcolor=blue]{hyperref}   
\usepackage[usenames,dvipsnames]{color}
\usepackage{amsfonts}
\usepackage{amsmath}
\usepackage{gensymb}
\usepackage{url}

\hypersetup{
    pdftitle = {Using electron vortex beams to determine chirality of crystals in transmission electron microscopy},
    pdfkeywords = {Vortex beam, electron vortex,electron diffraction, enantiomorph, chiral crystal}
}

\newcommand{\partieel}[2]{\ensuremath{\frac{\partial {#1}}{\partial {#2}}}}

\newcommand{\vt}[1]{\ensuremath{\boldsymbol{#1}}}
\newcommand{\sample}{Mn\ensuremath{_2}Sb\ensuremath{_2}O\ensuremath{_7}}

\newcommand{\braket}[3]{\ensuremath{\big<{#1}\big|{#2}\big|{#3}\big>}}
\newcommand{\Int}[1]{\ensuremath{\displaystyle\int \diff{#1}\,}}

\newcommand{\Intbep}[3]{\ensuremath{\displaystyle \int_{#1}^{#2}}\diff{#3}\,}
\newcommand{\diff}[2][]{\ifthenelse{\equal{#1}{}}{\ensuremath{\mathop{\mathrm{d} #2}}}{\ensuremath{\mathop{\mathrm{d} ^{#1} #2}}}}
\newcommand{\ket}[1]{\ensuremath{\left|{#1}\right>}}
\newcommand{\nextline}[1][]{\nonumber\\ 
#1&}

\begin{document}

\title{Using electron vortex beams to determine chirality of crystals in transmission electron microscopy}

\author{Roeland Juchtmans}
\affiliation{EMAT, University of Antwerp, Groenenborgerlaan 171, 2020 Antwerp, Belgium}
\author{Armand B\a'ech\a'e}
\affiliation{EMAT, University of Antwerp, Groenenborgerlaan 171, 2020 Antwerp, Belgium}
\author{Artem Abakumov}
\affiliation{EMAT, University of Antwerp, Groenenborgerlaan 171, 2020 Antwerp, Belgium}
\author{Maria Batuk}
\affiliation{EMAT, University of Antwerp, Groenenborgerlaan 171, 2020 Antwerp, Belgium}
\author{Jo Verbeeck}
\affiliation{EMAT, University of Antwerp, Groenenborgerlaan 171, 2020 Antwerp, Belgium}
\begin{abstract}
We investigate electron vortex beams elastically scattered on chiral crystals. After deriving a general expression for the scattering amplitude of a vortex electron, we study its diffraction on point scatterers arranged on a helix. We derive a relation between the handedness of the helix and the topological charge of the electron vortex on one hand, and the symmetry of the Higher Order Laue Zones in the diffraction pattern on the other for kinematically and dynamically scattered electrons. We then extend this to atoms arranged on a helix as found in crystals which belong to chiral space groups and propose a new method to determine the handedness of such crystals by looking at the symmetry of the diffraction pattern. Contrary to alternative methods, our technique does not require multiple scattering which makes it possible to also investigate extremely thin samples in which multiple scattering is suppressed. In order to verify the model, elastic scattering simulations are performed and an experimental demonstration on \sample\ is given where we find the sample to belong to the right handed variant of its enantiomorphic pair. This demonstrates the usefulness of electron vortex beams to reveal the chirality of crystals in a transmission electron microscope and provides the required theoretical basis for further developments in this field.\\
\end{abstract}
\maketitle

\section{Introduction}
An object is said to be chiral if it is not superposable onto its mirror image by rotating and/or translating it. As such, chiral objects come in two forms, an original and its mirror image called enantiomorphs. In crystals, chirality manifests itself in the crystal's space group (SG). These can be divided in three classes of which the first contains all the SGs having an improper symmetry operation (inversion center, mirror plane, glide plane or roto-inversion axes), which represent non-chiral crystals. Class II includes the 22 SGs which have a screw axis apart from the $2_1$-screw axis.  These SGs are chiral and can be divided in 11 enantiomorphic pairs, listed in table (\ref{TabEnantiomorphicSpaceGroups}). Because of the screw axis, atoms in these crystals are arranged in a helical way. The last class contains SGs which only possess proper rotations and/or a $2_1$-screw axis. Crystals belonging to these space groups also are chiral, but when mirrored the space group of the crystal doesn't change, therefore the SG itself is achiral.\\
\begin{table}[htb]
\centering
  \begin{tabular}[b]{r c r c r c}
  \hline
  \hline
  1.&($\mathrm{P}4_1, \mathrm{P}4_3$) & 5. & $(\mathrm{P}6_1, \mathrm{P}6_5)$ & 9. & $(\mathrm{P}3_1, \mathrm{P}3_2)$ \\
  2.& ($\mathrm{P}4_132, \mathrm{P}4_332$) & 6. & $(\mathrm{P}6_2, \mathrm{P}6_4)$ & 10. &$(\mathrm{P}3_112, \mathrm{P}3_212)$\\
  3.&  ($\mathrm{P}4_122, \mathrm{P}4_322$)& 7. & $(\mathrm{P}6_122, \mathrm{P}6_522)$ & 11. & $(\mathrm{P}3_121, \mathrm{P}3_221)$\\
  4.& ($\mathrm{P}4_12_12, \mathrm{P}4_32_12$) & 8. & $(\mathrm{P}6_222, \mathrm{P}6_422)$ &  &\\
  \hline
  \hline
  \end{tabular}
\caption{The 22 chiral space groups in class II, divided in 11 enantiomorphic space group pairs.}
\label{TabEnantiomorphicSpaceGroups}
\end{table}

Determining the chirality of a crystal is of importance when investigating its chiral properties, e.g. optical activity, chiral dichroism and chemical reactions with other chiral molecules. Doing this in a transmission electron microscope, however, appears to be a difficult task. In transmission electron microscopy, the crystal is seen (at least for the so-called zeroth order Laue zone (ZOLZ) in the diffraction pattern) as a two-dimensional projection through the thickness of the sample. Because of this, one can never distinguish between enantiomorphs since mirroring the crystal along the projection plane changes the handedness of the material, but not its projection on this plane. Therefore, all 3 dimensions have to be taken into account. In a first technique proposed by Goodman et al. \cite{Goodman1977a}, the chirality is recovered by rotating the object and taking several ZOLZ electron diffraction patterns. Another way to gather information on the dimension perpendicular to the projection plane, is to look at so-called Higher Order Laue Zones (HOLZ)\cite{Goodman1977}.
Friedel's law however imposes extra symmetry in the diffraction pattern that does not allow to determine the chirality in the kinematical approximation \cite{Friedel}. Therefore multiple scattering (dynamical approximation) is used to break this symmetry and to allow determining the chirality. All methods developed so far thus require dynamical simulations and depend in a sensitive way on sample thickness. An additional difficulty appears as not all HOLZ diffraction spots are sensitive to the handedness of the crystal and one has to identify them to be able to determine the chirality\cite{Johnson1994, Inui2003}. In this work, we propose the use of electron vortex beams to break the symmetry imposed by Friedel's law as an alternative way to distinguish SGs in one of the enantiomorphic SG pairs from table (\ref{TabEnantiomorphicSpaceGroups}) without having to rely on dynamical scattering.\\
As described theoretically by Nye and Berry \cite{Nye}, vortex waves are solutions of the 3D-wave equation with an angular dependent phase factor of the form
\begin{align}
\Psi(\vt{r})_m=\psi(r,z)e^{im\phi},
\end{align}
with $r$ and $\phi$ the radial and the azimuthal coordinate with respect to the wave propagation axis, $z$. The number $m$ is referred to as the topological charge. As an eigenstate of the angular momentum operator $L_z=-i\hbar\partieel{}{\phi}$, vortex waves carry a well-defined angular momentum of $m\hbar$ per photon or electron \cite{Allen}. In the center of the vortex, the phase is ill-defined and the intensity of the beam becomes zero due to destructive interference, resulting in the typical donut-like intensity profile. Ever since the first experimental realization of optical vortex beams over two decades ago \cite{Bazhenov1990}, they have been subject to active research which has led to applications varying from nano-manipulation \cite{Luo,He,Friese}, astrophysics \cite{Foo2005,Swartzlander2007,Serabyn2010,Berkhout2008} and telecommunication \cite{andrewsbook,wangterabit,vortexnot}. More recently, the first electron vortex beam \cite{Bliokh} was created by making a phase plate of stacked graphite layers \cite{Uchida2010}, followed by holographic reconstruction as a more reliable way of producing electron  vortices \cite{Verbeeck2010}. New ways are still being developed and currently single mode vortex beams at high current as well as atomic size vortices have been realized \cite{Beche2013, Clark2013}. Several studies suggest practical use of electron vortex beams in EMCD experiments \cite{Verbeeck2010}, nano-manipulation \cite{Verbeeck2013}, spin-polarization devices \cite{Karimi2012} and magnetic plasmons \cite{Mohammadi2012}.\\
As Friedel's law is only applicable to plane wave scattering, diffraction patterns obtained from vortex beams can be fundamentally different when scattered on enantiomorphic objects, even in the kinematical approximation. The central question in this paper is then: ``Are electron vortex beams capable of distinguishing chiral crystals within the kinematical approximation?'' We will tackle this problem by looking at crystals in chiral space groups in which the atoms are arranged along a helical axis. \\
\section{Theoretical formulation}
\subsection{Vortex scattering in cylindrical coordinates within the kinematical approximation \label{AppVortexScattering}}

Assuming the scattered part of the wave to be much smaller than its incoming part, the $1^{\text{st}}$ order Born-approximation considers the potential as a small perturbation on free space and describes only single scattering events. Within this approximation the scattering amplitude for an incoming wave $\psi_0(\vt{r})$ to scatter on a potential $V(\vt{r})$ to a plane wave with wave vector $\vt{k}'$ is proportional to \cite{DeGraef2003}

\begin{align}
A(\vt{k}')&=\braket{\vt{k}'}{V(\vt{r})}{\psi_0}\nonumber\\
&\propto\Int{\vt{r}} e^{-i\vt{k}'.\vt{r}}V(\vt{r})\psi_0(\vt{r}). \label{GeneralScatteringAmplitude}
\end{align}
Our aim is to find an expression for the vortex beam scattering amplitude, for which  $\psi_0(\vt{r})=\psi(r)e^{im\phi}e^{ik_zz}$. Starting from eq. \eqref{GeneralScatteringAmplitude}, we get for a vortex electron with topological charge $m$ on a potential $V(\vt{r})$:
\begin{align}
A_m(\vt{k}')&=\Int{\vt{r}} e^{-i\vt{k}'\cdot\vt{r}}V(\vt{r})e^{ik_z.z}e^{im\phi}\psi(r).\label{GeneralScatteringAmplitude}
\end{align}
Since vortex beams are most easily described in cylindrical coordinates, it will be convenient to expand the potential in a cylindrical symmetric basis of Bessel functions \cite{Wang}. 
\begin{align}
V(\vt{r})&=\frac{1}{\sqrt{2\pi}}\Int{k''_z}\Intbep{0}{\infty}{k''_\perp}k''_\perp\nonumber\\
&\sum_{m''=-\infty}^{\infty}V_{m''}(k''_\perp,k''_z)J_m''(k''_\perp r)e^{im''\phi}e^{ik''_zz},\label{AppExpandedPotential}
\end{align}
with
\begin{align}
V_{m''}(k''_\perp,k''_z)&=\frac{1}{\sqrt{2\pi}}\Int{z}\Intbep{0}{\infty}{r}\Intbep{0}{2\pi}{\phi}\nonumber\\
& \times rV(r,\phi,z)J_{m''}(k''_\perp r)e^{-im''\phi}e^{-i k''_zz}, \label{AppExpansionCoefficients}
\end{align}
Using these, we can write the scattering amplitude as
\begin{align}
A_m(\vt{k}')=&\Intbep{0}{\infty}{k''_\perp}k''_\perp \sum_{m'=-\infty}^{\infty}(-i)^{m'}V_{m'-m}(k''_\perp,k'_z-k_z)\nextline\times e^{im'\phi_{k'}}
\Intbep{0}{\infty}{r}rJ_{m'-m}(k''_\perp r)J_{m'}(k'_\perp r)\psi(r).\label{GeneralScattering}
\end{align}
Eq. \eqref{GeneralScattering} is the most general expression to discribe vortex scattering in cylindrical coordinates on a potential determined by its polar expansion coefficients $V_{m''}(k''_\perp,k''_z)$.

\subsection{Vortex scattering on a helix\label{AppHelixScattering}}

The potential of a helical distribution of point scatterers on a helix with pitch $P$, radius $R$ and $Q$ point scatterers on each period of the helix, can be written as
\begin{align}
V(\vt{r})&=\sum_{j,n=-\infty}^\infty\delta(r-R)\times\nonumber\\&\delta\left(\phi-a\frac{2\pi z}{P}+2\pi n\right)\delta\left(z-\frac{j}{Q}P\right) \label{AppDeltapotential}
\end{align}
where $a=+1$ in case of a right-handed and $a=-1$ for a left-handed helix. Starting from eq. \eqref{AppExpansionCoefficients}, we find the potential's expansion coefficients:

\begin{align}
V_{m,v}(q_\perp)=&\frac{RQ}{P}J_m(q_\perp R)\sum_N\delta\left(\frac{a m+v}{Q}-N\right),\label{DiscreteHelixCoefficients}
\end{align}
with $v\in \mathbb{Z}$ and $q_z=v\frac{2\pi}{P}$ is quantized because of the periodicity of the potential in the $z$-direction.\\
When dropping the constant $\frac{RQ}{P}$, filling the expansion coefficients \eqref{DiscreteHelixCoefficients} into eq. \eqref{GeneralScattering} and some simple algebra given in appendix, gives us the scattering amplitude
\begin{align}
A_m(\vt{k}')=& e^{i(m-a v)\phi_{k'}}\sum_{N=-\infty}^{\infty}(-i)^{NQ}e^{iNQ\phi_{k'}}\nonumber\\&\times J_{m-a v+NQ}(k'_\perp R)\psi(R),\label{AppScatteringHelix}
\end{align}
Note that, because of the periodicity in the $z$-direction, the transfered forward momentum, $k_z-k'_z=v\frac{2\pi}{P}$ is quantized. This means the diffraction pattern consists out of discrete rings which coincide with higher order Laue zones in conventional electron beam diffraction.\\
When looking at the scattering amplitude for certain rings for which $v=am+nQ$, or equivalently, $m-a v=nQ, n\in \mathbb{Z}$, we get
\begin{align}
&A_m({k}'_\perp,\phi_{k'},k'_z)|_{m- av=nQ}\nextline=J_{0}(k'_\perp R)  +\sum_{N=1}^{\infty}2(-i)^{NQ}J_{NQ}(k'_\perp R)\cos(NQ\phi_{k'}),\label{Sim}
\end{align}
for which
\begin{align}
A_m(k'_\perp,\phi_{k'},k'_z)|_{m-a v=nQ}&=A^*_m(k'_\perp,\phi_{k'},k'_z)|_{m-a v=nQ}\label{PointSymmetry2}
\end{align}
The intensity of the diffraction pattern, $I(\vt{k}')$ is given by the amplitude squared, $\left|A(\vt{k}')\right|^2$ and eq. \eqref{Sim} shows these diffraction rings will be centrosymmetric. Eq. \ref{PointSymmetry2} is the most interesting result in this paper. It means that the chirality of the helix can be seen from the symmetry of the HOLZs, at least for helices with odd-screw axis symmetry. Take for instance a right- or a left-handed 3-fold screw axis for which $a=+1$ resp. $a=-1$ and $Q=3$. When we look at the FOLZ ($v=1$) using a vortex beam with topological charge $m=1$, we see that the relation $1=v=a.m+nQ=a.1+n.3$ only is valid for $n=0$ and $a=1$. Thus the FOLZ of the right-handed helix will posses centrosymmetry, while that of the left-handed helix will not. 
\subsection{Effect of multiple scattering on symmetry of HOLZ}
\begin{figure}[h!]
\includegraphics[width=\columnwidth]{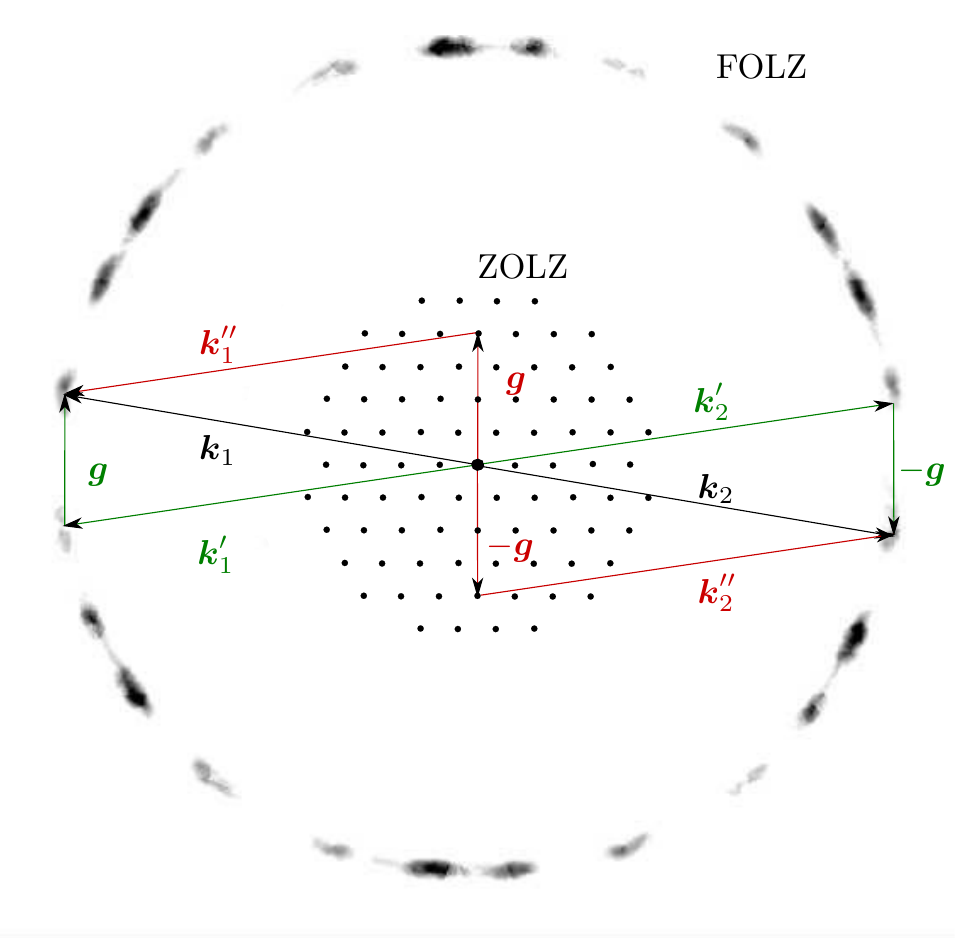}
\caption{[Colour online] Schematic diagram of the scattering paths of a focused vortex beam to scatter to the points $\vt{k}_1$ and $\vt{k}_2$, in the FOLZ opposite to each other. The black path represents the first order Born-approximation  in which the elctron scatters only once. The red path is a second order Born-approximation in which the electron first scatters to a point in the ZOLZ and then to the FOLZ. The green path represents an event in wich the electron scatters to the FOLZ and then scatters within this zone through a scattering vector of the ZOLZ. When neglecting multiple scattering between different Laue zones, the symmetry of the diffraction pattern is conserved also in the dynamical approximation.   \label{SecFOLZ}}
\end{figure}

The previous derivation of the symmetry of the diffraction rings was done in the kinematical approximation. However, electrons generally are subject to multiple scattering even when scattered on extremely thin samples. In this section we extend our findings on the symmetry of the FOLZ in the kinematical approximation to the situation where the electron is dynamically scattered.\\
Remember from the previous section that the scattering amplitude of a vortex electron with topological charge $m$ to scatter on a crystal to a plane wave with wave vector $\vt{k}$ is given by
\begin{align}
A(\vt{k})&=\braket{\vt{k}}{V(\vt{r})}{\psi_m}\nonumber\\
&=\Int{\vt{r}} e^{-i\vt{k}.\vt{r}}V(\vt{r})\psi_m(\vt{r}).\nonumber\\
&=\mathcal{F}[V(\vt{r})\psi_m(\vt{r})](-\vt{k})\nonumber\\
&=\mathcal{F}[V]*\mathcal{F}[\psi_m](-\vt{k}),
\end{align}
where $*$ denotes a convolution product.  \\
Let us now consider an electron that is subject to two scattering events while interacting with the potential. The scattering amplitude for the electron to scatter to a wave with wave vector $\vt{k}$ now is given by
\begin{align}
A^{(2)}_m(\vt{k})=&\Int{\vt{k}'}\braket{\vt{k}}{V(\vt{r})}{\vt{k}'}\braket{\vt{k}'}{V(\vt{r})}{\psi_m}.
\end{align}
It describes a scattering event of an electron in vortex state $\ket{\psi_m}$ to scatter to any plane wave $\ket{\vt{k}'}$ which is then scattered a second time to the final state $\ket{\vt{k}}$. We can rewrite this as
\begin{align}
A^{(2)}_m(\vt{k})=&\Int{\vt{k}'}\mathcal{F}[V](\vt{k}'-\vt{k})\left(\mathcal{F}[V]*\mathcal{F}[\psi_m](-\vt{k}')\right)\nonumber\\
=&\mathcal{F}[V]*\mathcal{F}[V]*\mathcal{F}[\psi_m](-\vt{k}).\label{SecondOrder}
\end{align}
When only considering scattering to and from the ZOLZ and FOLZ, eq. \eqref{SecondOrder} includes ZOLZ-ZOLZ, ZOLZ-FOLZ, FOLZ-ZOLZ and FOLZ-FOLZ scattering. The later contains two high angle scattering events between two different Laue zones and its contribution will be significantly smaller then the other events, which are schematically drawn in fig. (\ref{SecFOLZ}). Neglecting these, we show in App. \ref{AppDynamicScattering} that double scattering preserves the centrosymmetry we found before. 
\begin{align}
\left.A^{(2)}_m(k_\perp,\phi_k)\right|_{\vt{k}\in FOLZ}=
\left.A^{*(2)}_m(k_\perp,\phi_{k'}+\pi)\right|_{\vt{k}\in FOLZ}\label{DynamicSym}
\end{align}
 Moreover, this can easily be extended to dynamical scattering up to any order, when neglecting the scattering paths with more than one inter-Laue zone scattering event. We should note however that when more scattering events are taken into account, the relative contribution of scattering paths with more than one inter-Laue zone scattering event will increase, eventually breaking the symmetry of the FOLZ when looking at very thick samples.

\section{Electron vortex beams in a TEM}

\begin{figure}[htb]
\includegraphics[width=\columnwidth]{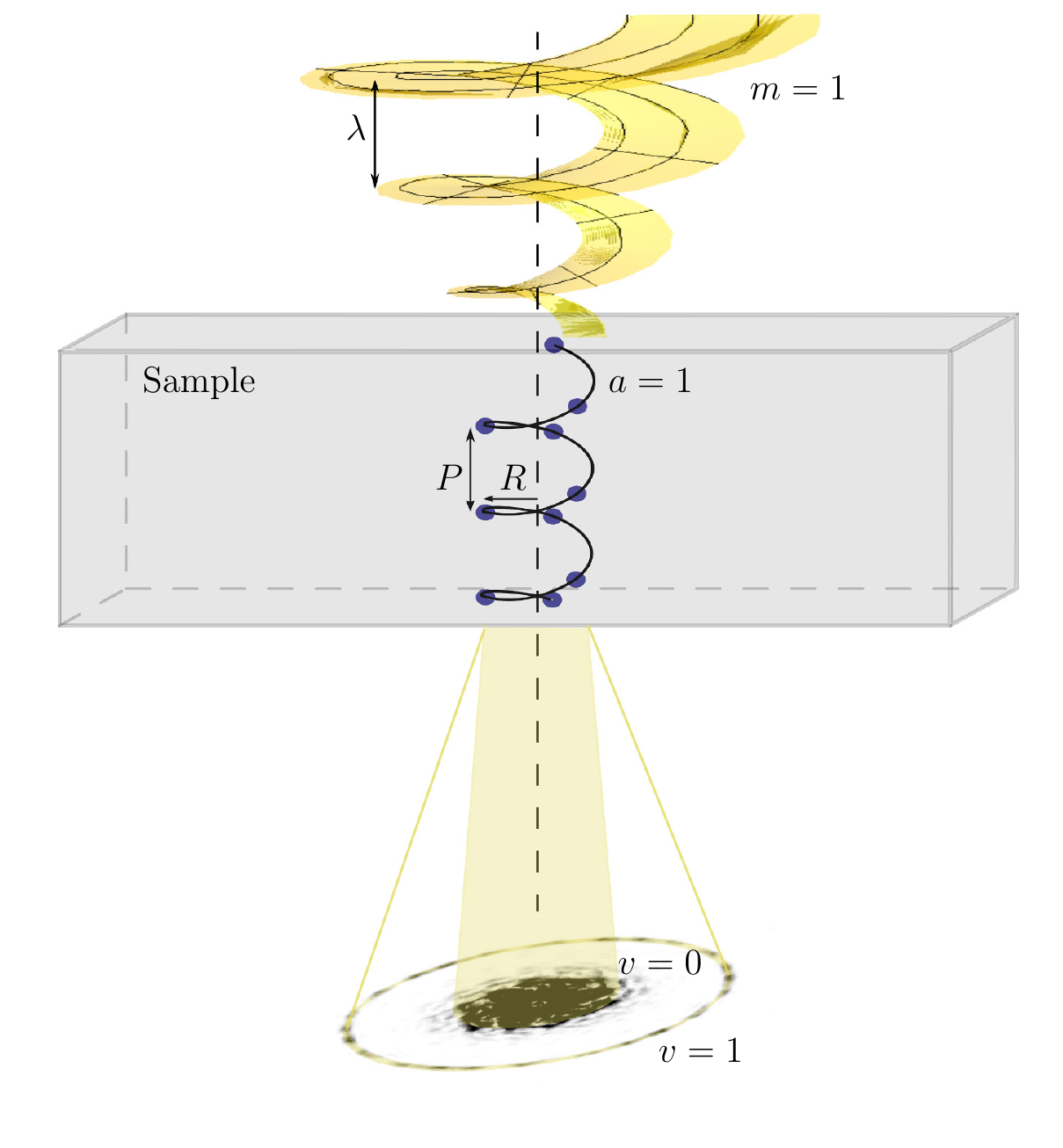}%
\caption{[Colour online] Schematic representation of the experimental setup. A vortex beam is focused directly on a screw axis in the sample further simplified assuming elastic scattering is dominated by the helically arranged heavy atoms. The radius of the vortex is chosen such that it matches the distance of the atoms to the screw axis. \label{ExperimentalSetup}}
\end{figure}

\begin{figure}[t!]
\centering
\includegraphics[width=\columnwidth]{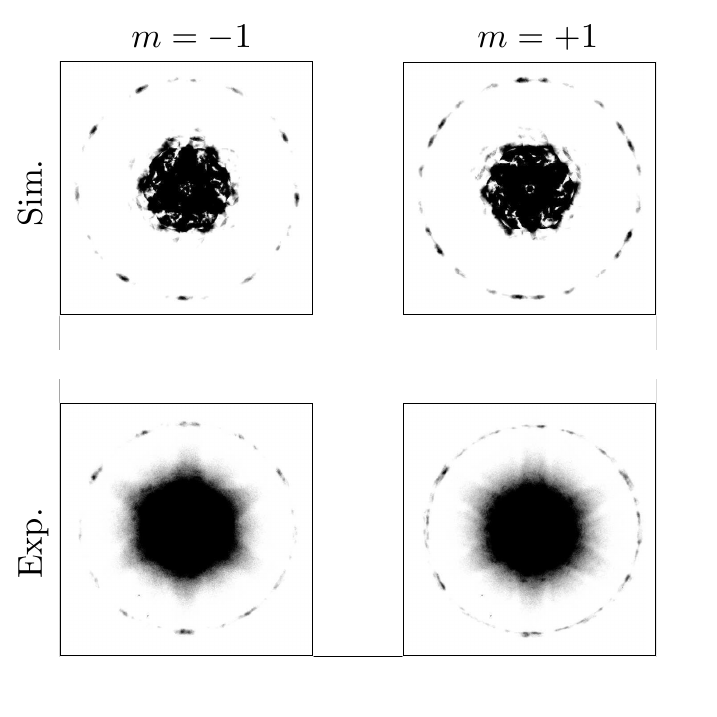}%
\caption{(Upper) Multislice simulation of  first order Laue zone of the diffraction pattern of a $m=-1$ (left) and an $m=+1$ vortex (right) scattered on right handed \sample, with SG $P3_121$. The vortex is centered along the screw axis with the size of the vortex equal to the distance of the Sb-atoms from the screw axis, approx. 1.2\,\AA. The energy of the probe is 300\,keV, the convergence angle 8\,mrad and the spherical aberration 1\,$\mu$m. The thickness of the sample is 20\,nm. The centrosymmetry of the first order ring can be seen for the right-handed enantiomorph for $m=+1$. (Lower) Experimental CBED pattern of a 1.2 \AA\ sized vortex probe pointed on a 3-fold screw axis in \sample\ for $m=-1$ (left) and $m=+1$ (right). The contrast was adapted to show only the peaks in the FOLZ.\label{FigCBED}}
\end{figure}

We apply the model of point scatterers on a helix as an approximation of a focused vortex beam scattered on a helical crystal when the vortex probe is centered on a screw axis and predominantly illuminates a set of heavy atoms distributed on a helix as sketched in fig. (\ref{ExperimentalSetup}). The crystal potential in real space can then be approximated by the convolution of the helix potential with the potential of a single heavy atom, while in Fourier space it is given by the product of the Fourier transform of the helix and the atomic potential. Since the latter is spherically symmetric, the angular dependency of the total Fourier transform stays the same. This means that, if we neglect the phase and intensity variation of the vortex beam over the size of one atom, the intensity profile of each Laue zone will still be given by eq. \eqref{AppScatteringHelix}. The only effect caused by the atomic scattering factor is a modulation of this intensity with respect to the scattering angle.\\
Eq. \eqref{AppScatteringHelix} shows that vortex beam diffraction is sensitive to the chirality of crystals. When a vortex beam is focussed on a screw axis with one heavy atom nearby, the HOLZs will look different for enantiomorphs. Moreover, for crystals with a 3-fold screw axis ($Q=3$), some of the HOLZs will become centrosymmetric depending on the topological charge of the vortex and the chirality of the crystal, revealing the latter at first glance.
In what follows we will demonstrate our new approach on \sample. From Scott et al. \cite{Scott1987} we know this crystal belongs to the space group $P3_121$ or $P3_221$ in which the atoms lie on a right- resp. left-handed 3-fold helix. One of the three inequivalent screw axes only has one heavy $Sb$-atom in its vicinity and electron scattering of a focused (vortex) electron beam centered on this axis approximates  the situation of scatterers on a helix. We will determine the chirality of the SG by determining the handedness of this screw axis, which we will refer to without further specification.\\
In order to verify whether the symmetry predicted in eq. \eqref{PointSymmetry2} still holds for a vortex beam scattered on a real crystal (i.e. not only a helical arrangement of atoms) we perform multislice simulations using the program STEMsim \cite{Rosenauer2007} where we include full dynamical scattering and spreading of the beam. We look at \sample\, along the crystallographic [001]-direction, the direction of the screw axis, and focus the vortex beam over it. We choose the semi-convergence angle of the beam such that the radius of the vortex matches the distance of the $Sb$-atoms to the screw axis, i.e. 1.2\,\AA. For a 300keV vortex beam with $m=\pm1$ the convergence angle corresponding to this criterion is 8\,mrad. The resulting CBED patterns for left- and right-handed vortices scattered on a right-handed 20\,nm thick \sample\ sample are given in fig. (\ref{FigCBED}). Here we see that the FOLZ of the left-handed vortex shows only 3-fold symmetry, whereas in the case of the right-handed vortex, the FOLZ shows sixfold symmetry.\\
Next we demonstrate our technique experimentally on crushed \sample\ crystals deposited on a carbon grid. The QuAnTem, a double $C_s$ corrected FEI Titan$^3$ microscope, operating at 300keV, was used to record the experimental diffraction patterns. The vortex beams were created with an aperture on which a magnetized needle is mounted, similar to the setup described by B\'ech\'e et al. \cite{Beche2013}.\\
After making a high-resolution image of a nano-crystal in the [001]-zone-axis, the beam was pointed exactly on the screw axis and the diffraction pattern was recorded with a CCD camera (1s exposure time). \\
We recorded five diffraction patterns using a vortex with topological charge $m\approx+1$ and six patterns with $m\approx-1$ scattered on the same crystal. Because the beam is much smaller than the unit cell size, the symmetry of the diffraction pattern is very sensitive to the position of the probe. Two images recorded with opposite vortices which showed the highest 3-fold symmetry in the diffraction pattern, are compared with the simulated ones in fig. (\ref{FigCBED}). We adjusted the contrast such that only the peaks in the FOLZ are visible and see that the recorded FOLZs qualitatively match the simulations for the right handed crystal rather well.\\
\begin{figure}[t!]
\centering
\includegraphics[width=\columnwidth]{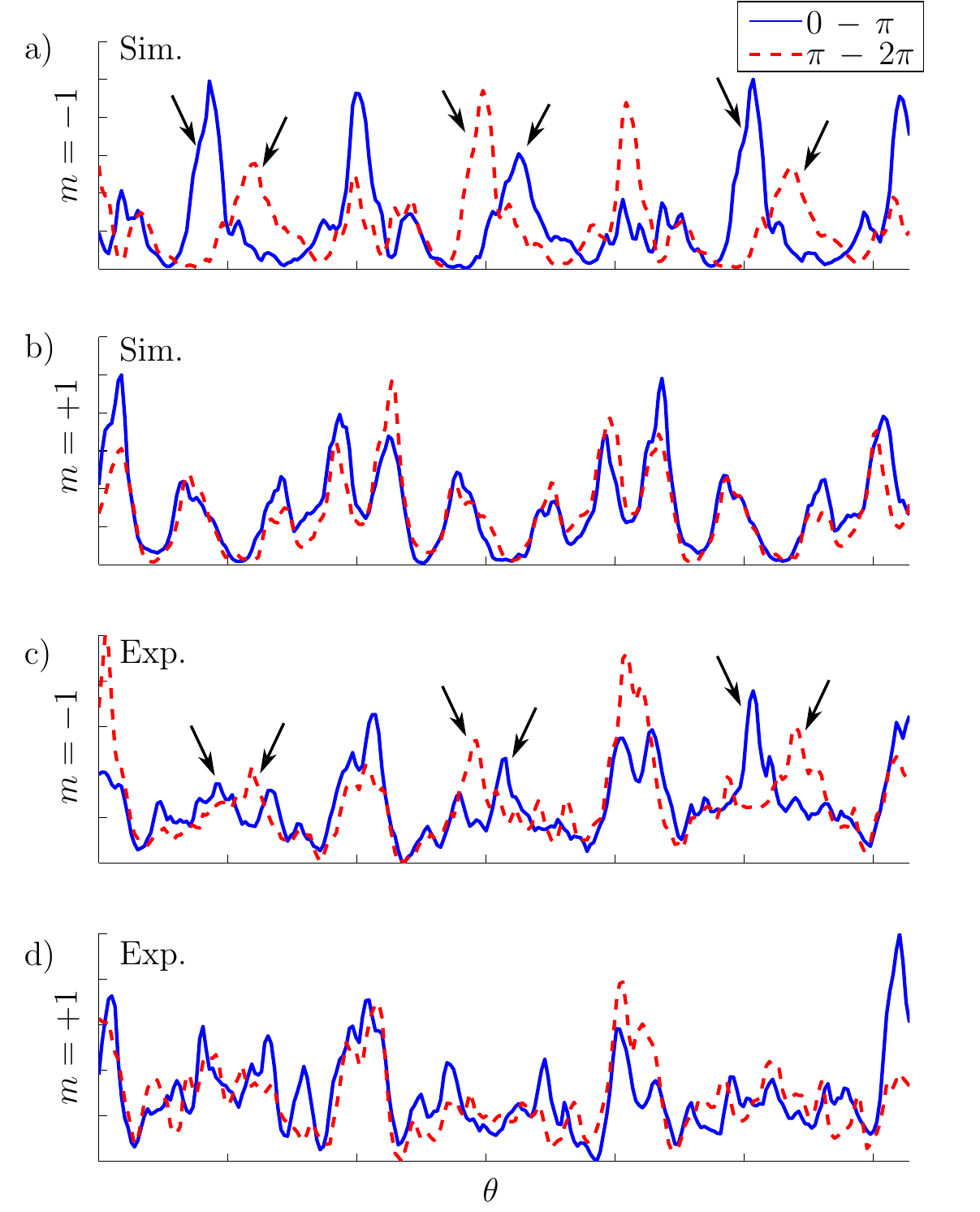}
\caption{[Colour online] Comparison of the simulated (a and b) and experimental (c and d) $I_{FOLZ}(\theta)$ for $\theta\in\{0,\pi\}$ (blue) and $\theta\in\{\pi,2\pi\}$ (red) for $m=-1$ (a and c) and $m=+1$ (b and d) on the $P3_121$ enantiomorph of \sample. From the position of the peaks it is clear that the $m=+1$ vortex scattered on the right-handed screw axis gives a centrosymmetric FOLZ, which is not the case for the $m=-1$ vortex, where the differences are indicated with arrows.\label{AngularSymmetry}}
\end{figure}
As it is our intention to determine the chirality of the crystal without comparing the experiment with simulation, we look at the symmetry of the FOLZ ring. To do so in more detail, we plot $I_{FOLZ}(\theta)$, the azimuthal intensity profile of the FOLZ integrated over a small radius $(r,r+d r)$, in fig. (\ref{AngularSymmetry}) for the azimuthal coordinate $\theta\in\{0,\pi\}$ (in blue) and for $\theta\in\{\pi,2\pi\}$ (in red). Although there are some differences between the $(0,\pi)$ and the $(\pi,2\pi)$-range for the simulated $m=+1$, the position of the peaks is clearly centrosymmetric, which is not at all the case for the simulated $m=-1$ (see arrows in fig. \ref{AngularSymmetry}). When looking at the experimental $I_{FOLZ}(\theta)$ it can be seen that most peaks in the $m=+1$ data have a peak at $\theta'=\theta+\pi$. For $m=-1$ we can see peaks in the $(0,\pi)$-range, which are not present in $(\pi,2\pi)$-range, indicating that the FOLZ here is not centrosymmetric. From this we conclude that the crystal under investigation has a right-handed screw axis and thus belongs to $P3_121$.

\begin{figure}[t!]
\includegraphics[width=\columnwidth]{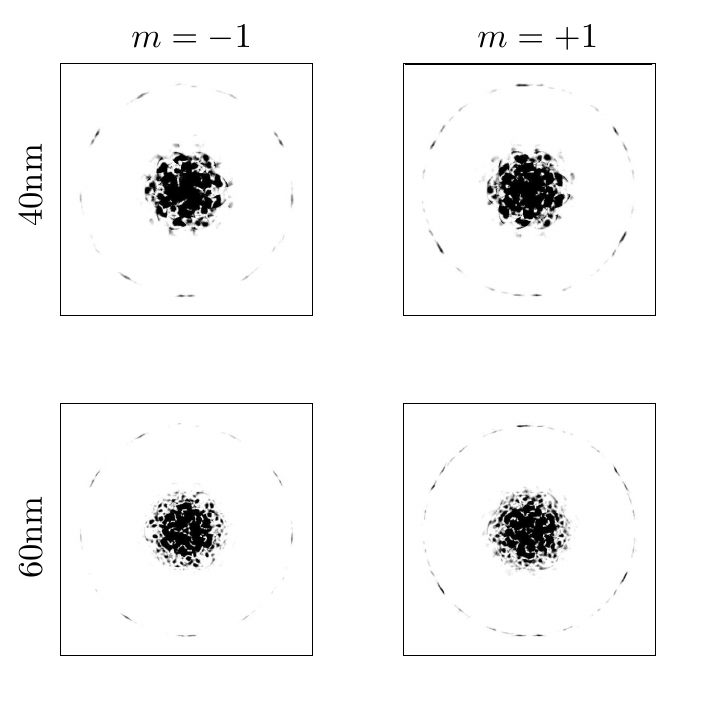}%
\caption{Multislice simulation of  first order Laue zone of the diffraction pattern of a $m=-1$ (left) and an $m=+1$ vortex (right) scattered on right handed \sample, with thickness 40\,nm (upper) and 60\,nm (lower). Although dynamical scattering is expected to be dominant, the centrosymmetry of the FOLZ still can be seen for $m=+1$, as was the case for the 20\,nm thick sample. \label{AppCBEDThickness}}
\end{figure}

\begin{figure}[h!]
\includegraphics[width=\columnwidth]{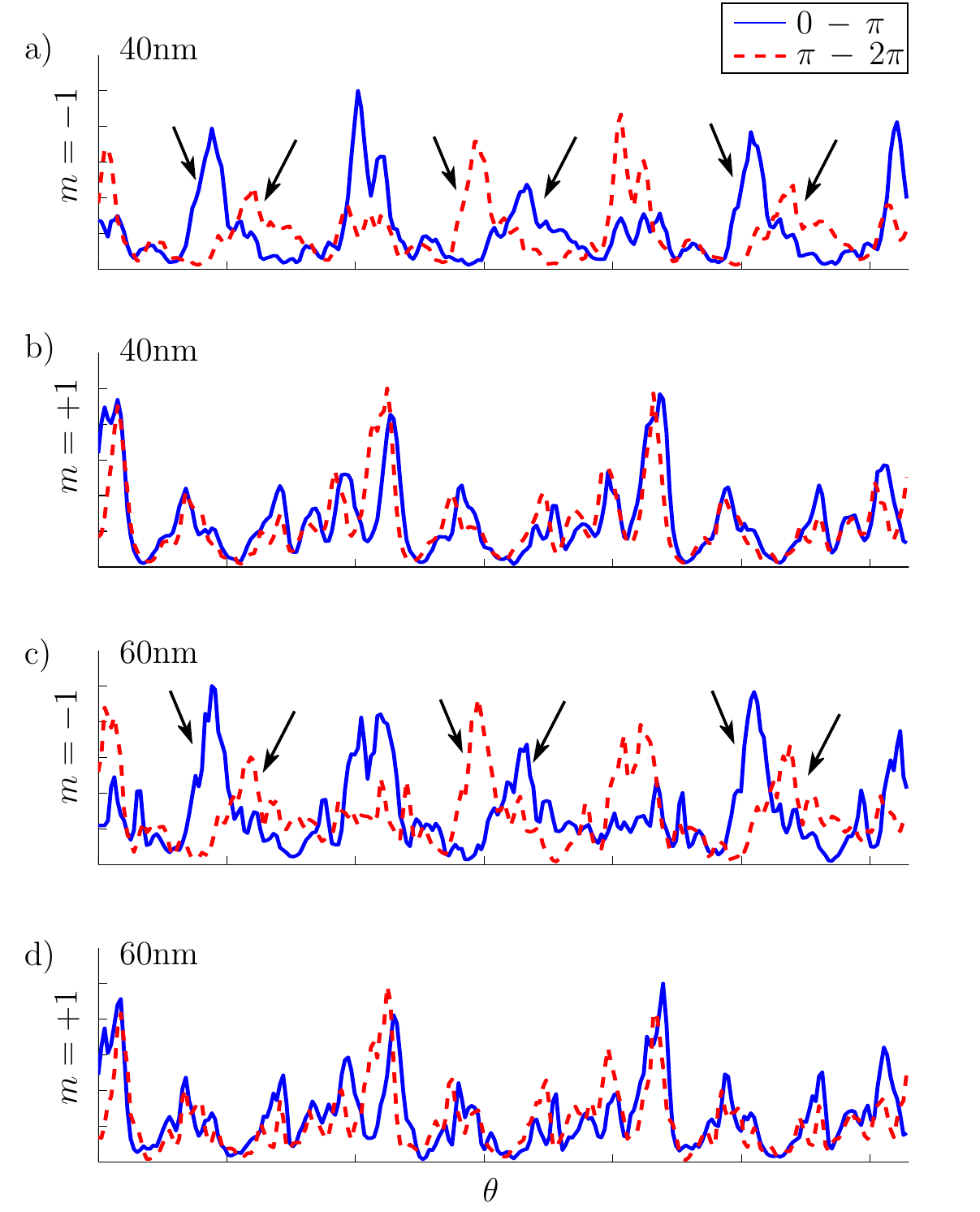}

\caption{[Colour online] Comparison of the simulated  $I_{FOLZ}(\theta)$ for $\theta\in\{0,\pi\}$ (blue) and $\theta\in\{\pi,2\pi\}$ (red) for $m=-1$ (a and c) and $m=+1$ (b and d) on the $P3_121$ enantiomorph of \sample\, for sample thicknesses 40\,nm (a and b) and 60\,nm (c and d). In b) and d) it can be seen that the positioning of the peaks still is centrosymmetric for samples up to 60\,nm, although the shape of the peaks gets altered. This symmetry is not at all present in the $m=-1$ vortex case where the main differences are indicated by arrows.\label{ThicknessProfiles}}
	\end{figure}

The major difficulty of this method and the main reason for the deviations of the symmetry in the experiment, is the probe positioning which has to be maintained during the acquisition of the diffraction pattern.\\
To demonstrate the robustness of the symmetry to multiple scattering we perform multi-slice simulation on sample thicknesses of 40\,nm and 60\,nm, shown in fig. (\ref{AppCBEDThickness}), where the dynamical scattering is expected to be dominant. When looking directly at the diffraction patterns, the centrosymmetry of the FOLZ of the $m=+1$ vortex appears to be conserved when scattered on the right handed crystal, even for a sample thickness up to 60\,nm. To look at this in more detail, we compare the intensity profile of the FOLZ for $\theta \in \{0,\pi\}$ with that for $\theta\in\{\pi,2\pi\}$, see fig. (\ref{ThicknessProfiles}). We can see that, although the shape of opposite peaks is different, the centrosymmetry again is present in the position of the peaks.

 \section{Conclusion}
We showed that vortex beam diffraction can be sensitive to the chirality of a crystal, even within the kinematical approximation. As an example, we calculated analytically the scattering amplitude of a vortex scattered kinematically on point scatterers arranged on a helix and found that the angular intensity profiles of the HOLZs depend on the topological charge of the vortex and the chirality of the helix. For helices with odd screw axis symmetry it was found that the chirality can be determined unambiguously from the symmetry in the diffraction pattern. We extend this property to dynamically scattered electrons that scatter only once from the ZOLZ to the FOLZ, increasing the range of its applicability to more conventional sample thicknesses in the TEM.\\
Based on these results, we demonstrate a new technique to determine the handedness  of a chiral SG by focusing  a vortex probe over a screw axis in the crystal. We verified the symmetry of the HOLZ with multislice simulations on \sample\ and give an experimental demonstration of the technique. The major difficulty is the required accuracy of the probe positioning, which, in combination with sample drift, makes acquisition of the data cumbersome. The 4D-STEM method proposed by Ophus et al. \cite{Ophus2014}  might help in  overcoming this problem. In this method a STEM image is recorded while taking a diffraction pattern at every scan point, thus avoiding the need to point the probe at the exact location manually.\\
Naturally, the use of vortices to determine the chirality of a crystal isn't restricted to crystals with a screw axis with only one heavy atom as nearest neighbor. In App. \ref{AppOtherMat} we show that in general, vortex wave diffraction is sensitive to the position of atoms in the direction perpendicular to the zone-axis within the kinematical approximation, making them a potentially valuable tool to determine the chirality of crystals, even when they don't belong to one of the chiral SGs in table \ref{TabEnantiomorphicSpaceGroups}. The symmetry of the diffraction pattern however, no longer can be used and simulations are needed to link the experiment with the chirality of the structure.\\
Although this technique has its experimental challenges, it has the advantage that the chirality of a crystal can be determined solely by looking at the symmetry of the diffraction pattern thus avoiding the need for dynamical calculations. Also, this can be done within the kinematical regime such that the chirality of very thin samples in which multiple scattering is limited, can be measured. On top of that, this can be done on a local basis, for each position in the crystal, opening the route for probing the chirality of crystals at atomic resolution or for investigating screw dislocations in non-chiral crystals. This work shows the first applications of electron vortex beam diffraction in crystallography where the chiral character of the probe has a significant influence on the diffraction pattern and its symmetry. It gives fundamental insights in how electron vortices interact with the symmetry of crystal and provides a basis for derived methods and techniques to be developed. Finally, it gives an example of the more general idea of optimizing the beam, both intensity and phase, to study specific properties, as suggested by Guzzinati et al. \cite{Guzzinati2014} and Rusz \cite{Rusz2014}.\\
The authors acknowledge support from the FWO (Aspirant Fonds Wetenschappelijk Onderzoek - Vlaanderen), the EU under the Seventh Framework Program (FP7) under a contract for an Integrated Infrastructure Initiative, Reference No. 312483-ESTEEM2, and the European Research Council under the FP7 and ERC Starting Grant 278510 VORTEX.
\nocite{DeGraef2003,Wang}

\appendix
\begin{widetext}
\section{Derivation of vortex scattering amplitude in cylindrical coordinates\label{AppScatteringAmplitude}}

Starting from the scattering amplitude within the first order Born-approximation for a vortex electron, eq.\ref{GeneralScatteringAmplitude} and writing the potential with its polar expansion coefficients, eq. \ref{AppExpandedPotential}, we get
\begin{align}
A_m(\vt{k}')=&\frac{1}{\sqrt{2\pi}}\Int{\vt{r}}\Int{k''_z}\Intbep{0}{\infty}{k''_\perp}k''_\perp \sum_{m''=-\infty}^{\infty}V_{m''}(k''_\perp,k''_z)\nonumber\\&J_{m''}(k''_\perp r)e^{im''\phi}e^{ik''_zz}e^{i(k_z-k'_z)z}e^{im\phi}e^{-i\vt{k}'_\perp\cdot\vt{r}_\perp}\psi(r).
\end{align}
This can be simplified using the Jacobi-Anger identity,
\begin{align}
e^{i\vt{k}_\perp\cdot\vt{r}}&=e^{ik_\perp r\cos(\phi-\phi_{k})}\nonumber\\
&=\sum_{m=-\infty}^{\infty}i^mJ_m(k_\perp r)e^{im(\phi-\phi_{k})},
\end{align}
to obtain
\begin{align}
A_m(\vt{k}')=&\Int{\vt{r}}\Int{k''_z}\Intbep{0}{\infty}{k''_\perp}k''_\perp\nonumber\\
\times& \sum_{m',m''=-\infty}^{\infty}(-i)^{m'}V_{m''}(k''_\perp,k''_z)J_{m''}(k''_\perp r)\nonumber\\\times& e^{i(m''+m-m')\phi}e^{i(k''_z+k_z-k'_z)z}
J_{m'}(k'_\perp r)e^{im'\phi_{k'}}\psi(r).
\end{align}
Where we omitted a factor $1/\sqrt{2\pi}$.
Performing the integrals over $\phi$ and $z$ gives us a Kronecker delta and a Dirac delta function, which finally gives us for the scattering amplitude:\\
\begin{align}
A_m(\vt{k}')=&\Intbep{0}{\infty}{r}r\Int{k''_z}\Intbep{0}{\infty}{k''_\perp}k''_\perp \sum_{m',m''=-\infty}^{\infty}(-i)^{m'}V_{m''}(k''_\perp,k''_z)J_{m''}(k''_\perp r)\delta_{m''+m-m'}\delta(k''_z+k_z-k'_z)
J_{m'}(k'_\perp r)e^{im'\phi_{k'}}\psi(r)\nonumber\\
=&\Intbep{0}{\infty}{k''_\perp}k''_\perp \sum_{m'=-\infty}^{\infty}(-i)^{m'}e^{im'\phi_{k'}}V_{m'-m}(k''_\perp,k'_z-k_z)
\Intbep{0}{\infty}{r}rJ_{m'-m}(k''_\perp r)J_{m'}(k'_\perp r)\psi(r).\label{AppGeneralScattering}
\end{align}

\section{Vortex scattering on a helix, derivation and properties\label{AppHelixScattering}}

Inserting the potential of point scatterers arranged on a helix, eq. \eqref{AppDeltapotential} , in the expression for the polar expansion coeffiecients, eq. \eqref{AppExpansionCoefficients} gives us
\begin{align}
V_{m}(q_\perp,q_z)=\Int{z}\Intbep{0}{\infty}{r}\Intbep{0}{2\pi}{\phi}rJ_m(q_\perp r)e^{-im\phi}e^{-i q_zz}
\sum_{j,n=-\infty}^\infty\delta(r-R)\delta\left(\phi-a\frac{2\pi z}{P}+2\pi n\right)\delta\left(z-\frac{j}{Q}P\right).\end{align}
We can rewrite this as
\begin{align}
V_{m,v}(q_\perp)=&\frac{1}{P}\Intbep{0}{P}{z}\Intbep{0}{2\pi}{\phi}RJ_m(q_\perp R)e^{-im\phi}e^{-i v\frac{2\pi}{P}z}
\sum_{j=1}^Q\delta\left(\phi-a\frac{2\pi z}{P}\right)\delta\left(z-\frac{j}{Q}P\right)\nonumber\\
=&\frac{R}{P}J_m(q_\perp R)\sum_{j=1}^Qe^{-i(a m+v)\frac{2\pi j}{Q}}\nonumber\\
=&\frac{RQ}{P}J_m(q_\perp R)\sum_N\delta\left(\frac{a m+v}{Q}-N\right),\label{AppDiscreteHelixCoefficients}
\end{align}
with $v\in \mathbb{Z}$ and $q_z=v\frac{2\pi}{P}$. The latter is quantized because of the periodicity in $z$.\\
When dropping the constant $\frac{RQ}{P}$, filling the expansion coefficients \eqref{AppDiscreteHelixCoefficients} into eq. \eqref{AppGeneralScattering} gives us for the scattering amplitude
\begin{align}
A_m(\vt{k}')=&\Intbep{0}{\infty}{k''_\perp}k''_\perp \sum_{N,m'=-\infty}^{\infty}(-i)^{m'}e^{im'\phi_{k'}} J_{m'-m}(k_\perp'' R)
\delta\left(\frac{a(m'-m)+v}{Q}-N\right)\nonumber\\&\times\Intbep{0}{\infty}{r}rJ_{m'-m}(k''_\perp r)J_{m'}(k'_\perp r)\psi(r).
\end{align}
where the number $v=\frac{P}{2\pi}(k_z-k'_z)\in \mathbb{Z}$ discretizes the transfered forward momentum.
The integral over $k"_\perp$ is given analytically by
\begin{align}
\Intbep{0}{\infty}{k''_\perp}k''_\perp J_{n}(k''_\perp r)J_{n}(k''_\perp R)=\frac{\delta(r-R)}{r},
\end{align}
 such that we find the final scattering amplitude to be
\begin{align}
A_m(\vt{k}')=& \sum_{N,m'=-\infty}^{\infty}(-i)^{m'}e^{im'\phi_{k'}}\delta\left(\frac{a(m'-m)+v}{Q}-N\right)
\Intbep{0}{\infty}{r}\delta(r-R)J_{m'}(k'_\perp r)\psi(r)\nonumber\\
=& e^{i(m-av)\phi_{k'}}\sum_{N=-\infty}^{\infty}(-i)^{a NQ}e^{iaNQ\phi_{k'}}\times J_{m+a(NQ-v)}(k'_\perp R)\psi(R)\nonumber\\
=& e^{i(m-a v)\phi_{k'}}\sum_{N=-\infty}^{\infty}(-i)^{NQ}e^{iNQ\phi_{k'}}\times J_{m-a v+NQ}(k'_\perp R)\psi(R),\label{AppScatteringAmplitudeHelix}
\end{align}

When looking at the scattering amplitude for rings for which $v=am+nQ$, or equivalently, $m-a v=nQ, n\in \mathbb{Z}$, we get
\begin{align}
A_m(k'_\perp,\phi_{k'})|_{m- av=nQ}=&e^{inQ\phi_{k'}}\sum_{N=-\infty}^{\infty}(-i)^{NQ}e^{iNQ\phi_{k'}}J_{nQ+NQ}(k'_\perp R)\nonumber\\
=&i^{nQ}\sum_{N=-\infty}^{\infty}(-i)^{NQ}e^{iNQ\phi_{k'}}J_{NQ}(k'_\perp R)\nonumber\\
=&J_{0}(k'_\perp R)+ \sum_{N=1}^{\infty}\left((-i)^{NQ}e^{iNQ\phi_{k'}}J_{NQ}(k'_\perp R) +(-i)^{-NQ}e^{-iNQ\phi_{k'}}J_{-NQ}(k'_\perp R) \right)\nonumber\\
=&J_{0}(k'_\perp R)+ \sum_{N=1}^{\infty}(-i)^{NQ}J_{NQ}(k'_\perp R)\left(e^{iNQ\phi_{k'}} +e^{-iNQ\phi_{k'}}\right)\nonumber\\
=&J_{0}(k'_\perp R)+ \sum_{N=1}^{\infty}2(-i)^{NQ}J_{NQ}(k'_\perp R)\cos(NQ\phi_{k'}).\label{AppMAV}
\end{align}\\
Where we omitted a phase factor $i^{nQ}$. Looking at the scattering amplitude in the point $(k'_\perp,\phi_{k'}+\pi)$, gives us\\
\begin{align}
A_m(k'_\perp,\phi_{k'}+\pi)|_{m-a v=nQ}&=J_{0}(k'_\perp R)+ \sum_{N=1}^{\infty}2(-i)^{NQ}J_{NQ}(k'_\perp R)\cos(NQ(\phi_{k'}+\pi)) \nonumber\\
&=J_{0}(k'_\perp R)+ \sum_{N=1}^{\infty}2(i)^{NQ}J_{NQ}(k'_\perp R)\cos(NQ(\phi_{k'})) \nonumber\\
&=A^*_m(k'_\perp,\phi_{k'})|_{m-a v=nQ}\label{AppPointSymmetry2}
\end{align}

\section{Effect of multiple scattering on symmetry of HOLZ, derivations\label{AppDynamicScattering} }

As the crystal potential is periodic, its Fourier transform is a discrete function which is only non-zero at the reciprocal lattice points
\begin{align}
\mathcal{F}[V](\vt{k})=\sum_{\vt{g}}V_{\vt{g}}\delta(\vt{k}-\vt{g}),
\end{align}
with $\vt{g}$ a reciprocal lattice vector. As a consequence of this, a plane wave propagating through a crystal, can only scatter to another wave of which the wavevector differs by a reciprocal wave vector, $\vt{k}'\rightarrow\vt{k}=\vt{k}'+\vt{g}$. When we only take second order scattering events between the ZOLZ, the FOLZ  and the $FOLZ^*$ ($\Delta k_z<0$) into account and neglect higher order Laue zones contributions, we can write \eqref{SecondOrder} as
\begin{align}
A^{(2)}_m(\vt{k})=&\left(\sum_{\vt{g}\in ZOLZ}V_{\vt{g}}\delta(\vt{k}+\vt{g})+\sum_{\vt{g}\in FOLZ}V_{\vt{g}}\delta(\vt{k}+\vt{g})+\sum_{\vt{g}\in FOLZ^*}V_{\vt{g}}\delta(\vt{k}+\vt{g})\right)\nonumber\\
&*\left(\sum_{\vt{g}'\in ZOLZ}V_{\vt{g}'}\delta(\vt{k}+\vt{g}')+\sum_{\vt{g}'\in FOLZ}V_{\vt{g}'}\delta(\vt{k}+\vt{g}')+\sum_{\vt{g}'\in FOLZ^*}V_{\vt{g}'}\delta(\vt{k}+\vt{g}')\right)*\mathcal{F}[\psi_m](\vt{-k})\nonumber\\
=&\left(\left(\sum_{\vt{g}\in ZOLZ}V_{\vt{g}}\delta(\vt{k}+\vt{g})*\sum_{\vt{g}'\in ZOLZ}V_{\vt{g}'}\delta(\vt{k}+\vt{g}')\right)+\left(\sum_{\vt{g}\in ZOLZ}V_{\vt{g}}\delta(\vt{k}+\vt{g})*\sum_{\vt{g}'\in FOLZ}V_{\vt{g}'}\delta(\vt{k}+\vt{g}')\right)\right.\nonumber\\&+\left(\sum_{\vt{g}\in FOLZ}V_{\vt{g}}\delta(\vt{k}+\vt{g})*\sum_{\vt{g}'\in ZOLZ}V_{\vt{g}'}\delta(\vt{k}+\vt{g}')\right)+\left(\sum_{\vt{g}\in FOLZ}V_{\vt{g}}\delta(\vt{k}+\vt{g})*\sum_{\vt{g}'\in FOLZ^*}V_{\vt{g}'}\delta(\vt{k}+\vt{g}')\right)\nonumber\\&\left.+\left(\sum_{\vt{g}\in FOLZ^*}V_{\vt{g}}\delta(\vt{k}+\vt{g})*\sum_{\vt{g}'\in FOLZ}V_{\vt{g}'}\delta(\vt{k}+\vt{g}')\right)+\left(\sum_{\vt{g}\in FOLZ}V_{\vt{g}}\delta(\vt{k}+\vt{g})*\sum_{\vt{g}'\in FOLZ}V_{\vt{g}'}\delta(\vt{k}+\vt{g}')\right)\right)\nonumber\\&*\mathcal{F}[\psi_m](\vt{-k}),\label{DoubleScattering}
\end{align}
where we dropped the contributions of $ZOLZ$-$FOLZ^*$ and $FOLZ^*$-$FOLZ^*$  scattering events since these are energetically forbidden. At this point, we can make a second approximation in which we drop all the terms involving scattering through two reciprocal lattice points in the $FOLZ$ and $FOLZ^*$. These are associated with two high angle scattering events and therefore are fare less likely to occur compared to the scattering events within the ZOLZ. What remains is
\begin{align}
A^{(2)}_m(\vt{k})
=&\left(\left(\sum_{\vt{g}\in ZOLZ}V_{\vt{g}}\delta(\vt{k}+\vt{g})*\sum_{\vt{g}'\in ZOLZ}V_{\vt{g}'}\delta(\vt{k}+\vt{g}')\right)+2\left(\sum_{\vt{g}\in ZOLZ}V_{\vt{g}}\delta(\vt{k}+\vt{g})*\sum_{\vt{g}'\in FOLZ}V_{\vt{g}'}\delta(\vt{k}+\vt{g}')\right)\right)\nonumber\\&*\mathcal{F}[\psi_m](\vt{-k})
\end{align}
where we made use of the associativity of the convolution product, $A* B=B* A$. The first term describes two scattering events within the ZOLZ that don't alter the intensity of the FOLZ. So in order to study the symmetry of the FOLZ, we only have to look at the second term. 
\begin{align}
\left.A^{(2)}_m(\vt{k})\right|_{\vt{k}\in FOLZ}
\propto&\sum_{\vt{g}\in ZOLZ}V_{\vt{g}}\delta(\vt{k}+\vt{g})*\sum_{\vt{g}'\in FOLZ}V_{\vt{g}'}\delta(\vt{k}+\vt{g}')*\mathcal{F}[\psi_m](\vt{-k})\nonumber\\
\propto&\sum_{\vt{g}\in ZOLZ}V_{\vt{g}}\delta(\vt{k}+\vt{g})* \left.A^{(1)}_m(\vt{k})\right|_{\vt{k}\in FOLZ}\label{38}
\end{align}
where $A^{(1)}_m(\vt{k})$ is the scattering amplitude within the first order Born-approximation of our focused vortex beam from the previous section. Depending on the chirality of the helix and the topological charge of the vortex, we know that this can be a centrosymmetric function or not, that is when we scatter a right handed vortex on a right handed helix, we get \begin{align}
 A^{(1)}_m\left(\vt{k}_\perp,\frac{2\pi}{P}\right)&=A^{*(1)}_m\left(-\vt{k}_\perp,\frac{2\pi}{P}\right)
\end{align}
Also for a crystal with real potential (elastic scattering), we get from Friedel's law
\begin{align}
V_{\vt{g}}=V^*_{-\vt{g}}
\end{align}
The convolution product in eq. \eqref{38} now yields
\begin{align}
\left.A^{(2)}_m(\vt{k})\right|_{\vt{k}\in FOLZ}\propto&\Int{\vt{k}'}\sum_{\vt{g}\in ZOLZ}V_{\vt{g}}\delta(-\vt{k}-\vt{k}'-\vt{g}) \left.A^{(1)}_m(\vt{k}')\right|_{\vt{k}'\in FOLZ}\nextline[\propto]
\Int{k_z'}\delta(k_z-k_z')\Int{\vt{k}'_\perp}\sum_{\vt{g_\perp}}V_{\vt{g}_\perp}\delta(-\vt{k}_\perp-\vt{k}'_\perp-\vt{g}_\perp) \left.A^{(1)}_m(\vt{k}'_\perp,k'_z)\right|_{\vt{k}'\in FOLZ}\nextline[\propto]
\Int{k_z'}\delta(k_z-k_z')\Int{(-\vt{k}'_\perp)}\sum_{\vt{g_\perp}}V_{-\vt{g}_\perp}\delta(-\vt{k}_\perp+\vt{k}'_\perp+\vt{g}_\perp) \left.A^{(1)}_m(-\vt{k}'_\perp,k'_z)\right|_{\vt{k}'\in FOLZ}\nextline[\propto]
\Int{k_z'}\delta(k_z-k_z')\Int{\vt{k}'_\perp}\sum_{\vt{g_\perp}}V^*_{\vt{g}_\perp}\delta(+\vt{k}_\perp-\vt{k}'_\perp-\vt{g}_\perp) \left.A^{*(1)}_m(\vt{k}'_\perp,k'_z)\right|_{\vt{k}'\in FOLZ}\nextline[\propto]
\left.A^{*(2)}_m(-\vt{k}_\perp,k_z)\right|_{\vt{k}\in FOLZ}\label{DynamicSym}
\end{align}
Taking into account double scattering events that include only one high angle scattering event between ZOLZ and FOLZ, schematically illustrated in fig. (\ref{SecFOLZ}), preserves the centrosymmetry we found before. Moreover, this can easily be extended to dynamical scattering up to any order, when neglecting the scattering paths with more than one inter-Laue zone scattering event. Suppose the $n^{\text{th}}$-order Born-approximation for the scattering amplitude $A^{(n)}_m(\vt{k})|_{\vt{k}\in FOLZ}$ is centrosymmetric. The $(n+1)^{\text{th}}$-order is then given by
\begin{align}
&\left.A^{(n+1)}_m(\vt{k})\right|_{\vt{k}\in FOLZ}\nonumber\\ &\propto\Int{\vt{k}'}\sum_{\vt{g}\in ZOLZ}V_{\vt{g}}\delta(-\vt{k}-\vt{k}'-\vt{g}) \left.A^{(n)}_m(\vt{k}')\right|_{\vt{k}'\in FOLZ}\nonumber\\&\propto
\left.A^{*(n+1)}_m(-\vt{k}_\perp,k_z)\right|_{\vt{k}\in FOLZ},
\end{align}

\section{What about other chiral potentials?\label{AppOtherMat}}
When one takes a 2D image of a potential, one can never distinguish between two enantiomorphs since mirroring the crystal along the image plane would change the crystal's potential, but not the 2D image. HOLZs in the diffraction pattern on the other hand, carry information about the crystal in the direction parallel to the direction of view and one can think of using these to determine the chirality. However, within the kinematical approximation, this is not possible using a conventional probe of which the phase is constant in any plane parallel to the image plane.\\
Take for instance a potential $V^R(\vt{r})$ and its enantiomorph $V^L(\vt{r})=V^R(x,y,-z)$, obtained by mirroring the potential along the plane of view. Their polar expansion coefficients are related by
\begin{align}
V^R_m(k\perp,k_z)=&\Int{z}\Intbep{0}{\infty}{r}\Intbep{0}{2\pi}{\phi}\times rV^R(r,\phi,z)J_m(k_\perp r) e^{-im\phi}e^{-ik_zz}\nextline[=]
\Int{z}\Intbep{0}{\infty}{r}\Intbep{0}{2\pi}{\phi}  \times rV^L(r,\phi,-z)J_{m}(k_\perp r) e^{-im\phi}e^{-ik_zz}\nextline[=]
\Int{z}\Intbep{0}{\infty}{r}\Intbep{0}{2\pi}{\phi}  \times rV^L(r,\phi,z)(-1)^mJ_{-m}(k_\perp r) e^{i(-m)\phi}e^{ik_zz}\nextline[=]
(-1)^mV^{*L}_{-m}(k_\perp,k_z).
\end{align}
As a consequence of this, the diffraction patterns, $\left|A_0(\vt{k}')\right|^2$ only differ by a rotation of $\pi$\,rad.

\begin{align}
A^R_0(\vt{k}')=&\Intbep{0}{\infty}{k''_\perp}k''_\perp \sum_{m'=-\infty}^{\infty}(-i)^{m'} V^R_{m'}(k''_\perp,k'_z-k_z) e^{im'\phi_{k'}}\Intbep{0}{\infty}{r}rJ_{m'}(k''_\perp r)J_{m'}(k'_\perp r)\psi(r)\nextline[=]
\Intbep{0}{\infty}{k''_\perp}k''_\perp \sum_{m'=-\infty}^{\infty}(-i)^{m'}V^{*L}_{-m'}(k''_\perp,k'_z-k_z)  e^{im'(\phi_{k'}+\pi)} \Intbep{0}{\infty}{r}rJ_{m'}(k''_\perp r)J_{m'}(k'_\perp r)\psi(r)\nextline[=]
\Intbep{0}{\infty}{k''_\perp}k''_\perp \sum_{m'=-\infty}^{\infty}(-i)^{-m'}V^{*L}_{m'}(k''_\perp,k'_z-k_z) e^{-im'(\phi_{k'}+\pi)} \Intbep{0}{\infty}{r}rJ_{-m'}(k''_\perp r)J_{-m'}(k'_\perp r)\psi(r)\nextline[=]
A^{*L}_0(k'_\perp,\phi_{k'}+\pi,k'_z),\label{AppPlaneWaveScattering}
\end{align}
where in the next to last step we made use of $J_{m'}(k''_\perp r)J_{m'}(k'_\perp r)=(-1)^{m'+m'}J_{-m'}(k''_\perp r)J_{-m'}(k'_\perp r)$ and in the last step we changed $-m'\rightarrow m'$ since the summation goes over all possible $m'$. The Bessel functions in eq. \eqref{AppPlaneWaveScattering} give the same weight to expansion coefficients $V_m$ and $V_{-m}$ to the diffraction pattern and it is this property that makes enantiomorphs indistinguishable by using plane waves. \\
When considering scattering of a vortex beam on the other hand, in eq. \eqref{AppGeneralScattering} we have the term
\begin{align}
J_{m'+m}(k''_\perp r)J_{m'}(k'_\perp r)\neq J_{-m'+m}(k''_\perp r)J_{-m'}(k'_\perp r).
\end{align}
\end{widetext}
This term gives a different weight to the contribution of the expansion coefficients $V_m$ and $V_{-m}$ to the diffraction pattern and therefore the diffraction patterns of enantiomorphic potentials in general will look different when using vortex beams even within the kinematical approximation.\\
As an example we simulated the diffraction pattern of a chiral set of point scatterers with 4-fold symmetry and periodicity in the $z$-direction. The beam is pointed exactly on the 4-fold rotation axis making the diffraction pattern 4-fold symmetric. Because of the periodicity along the beam direction, the diffraction pattern consists out of rings corresponding with ZOLZ and HOLZs. As a basis we took three points with coordinates (in \AA) $C_1=(0.5,0,0)$, $C_2=(0.5,0.5,0)$ and $C_3=(0.5,0,\pm0.25)$ for resp. the right handed and left handed enantiomorph. This basis was rotated around the origin over angles $\pi/2$\,rad, $\pi$\,rad and $3\pi/2$\,rad and repeated 10 times in the $z$-direction with period 1\,\AA. These points were considered Huygens point sources emitting waves with a wavelength corresponding to a 300\,keV electron beam of which the phase depends on the $z$-direction and, in case of a vortex probe, on the angular coordinate.  The resulting diffraction pattern is shown in fig. (\ref{4FoldSim}). We see that, within the kinematical approximation, the plane wave diffraction pattern is identical for the left or the right handed enantiomorph. When we use a vortex probe, $m=1$ on the other hand, we clearly see a difference between the two diffraction patterns. This demonstrates that vortex beam scattering in general is sensitive to the chirality of a potential, not only to the specific case we investigated in the article.
\begin{figure}[h!]
\includegraphics[width=\columnwidth]{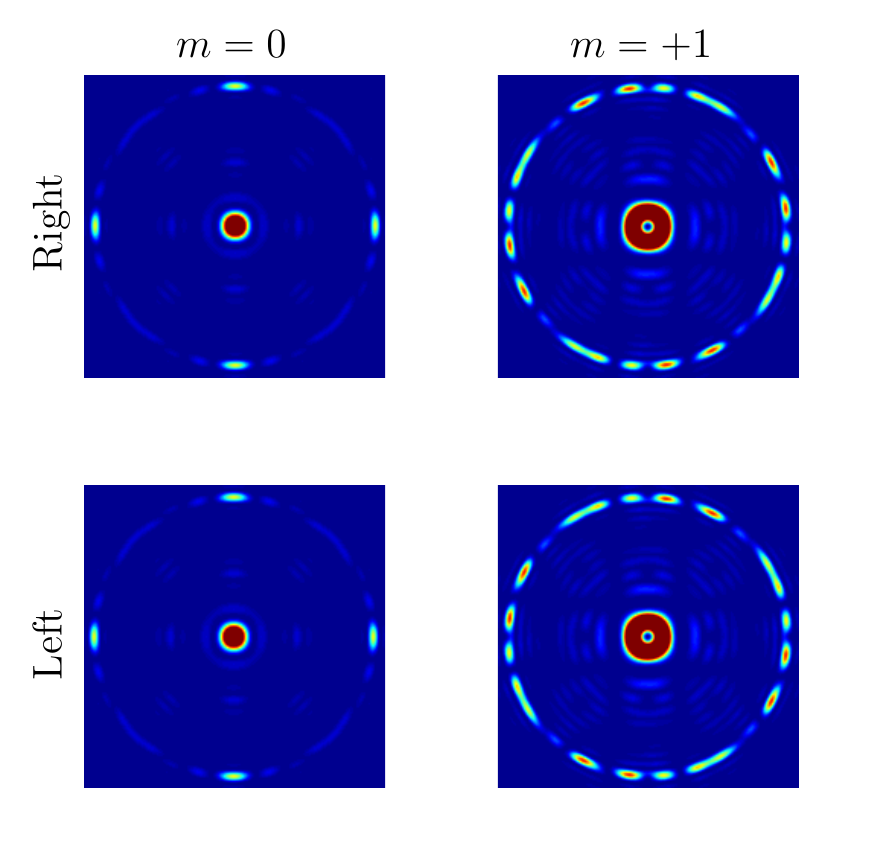}
\caption{[Colour online] Simulated diffraction pattern of a chiral set of point scatterers with 4-fold symmetry for a plane wave and a vortex probe. When the right handed enantiomorph is mirrored along the beam direction to get the left handed enantiomorph, the diffraction of the plane wave stays identical, while that of the vortex is different. This shows that vortex beam scattering in the kinematical approximation will, in contrast to plane wave scattering, depend on the chirality of a potential. \label{4FoldSim}}
\end{figure}

\end{document}